\documentclass[conference]{IEEEtran}
\IEEEoverridecommandlockouts
\usepackage{cite}
\usepackage{amsmath,amssymb,amsfonts}
\usepackage{algorithmic}
\usepackage{hyperref}
\usepackage{graphicx}
\usepackage{textcomp}
\usepackage{xcolor}
\usepackage{bm}
\usepackage{cite}
\def\BibTeX{{\rm B\kern-.05em{\sc i\kern-.025em b}\kern-.08em
    T\kern-.1667em\lower.7ex\hbox{E}\kern-.125emX}}
\begin{document}

\title{Adiabatic Quantum Optimization Fails to Solve the Knapsack Problem\\
\thanks{This manuscript has been authored in part by UT-Battelle, LLC under Contract No. DE-AC05-00OR22725 with the U.S. Department of Energy. The United States Government retains and the publisher, by accepting the article for publication, acknowledges that the United States Government retains a non-exclusive, paid-up, irrevocable, world-wide license to publish or reproduce the published form of this manuscript, or allow others to do so, for United States Government purposes. The Department of Energy will provide public access to these results of federally sponsored research in accordance with the DOE Public Access Plan (http://energy.gov/downloads/doe-public-access-plan). This research used resources of the Oak Ridge Leadership Computing Facility, which is a DOE Office of Science User Facility supported under Contract DE-AC05-00OR22725.}
}

\author{
\IEEEauthorblockN{Lauren Pusey-Nazzaro}
\IEEEauthorblockA{\textit{Department of Physics} \\
\textit{Washington University}\\
St. Louis, MO \\
lauren.p@wustl.edu}
\and
\IEEEauthorblockN{Prasanna Date}
\IEEEauthorblockA{\textit{Computer Science and Mathematics} \\
\textit{Oak Ridge National Laboratory}\\
Oak Ridge, TN \\
datepa@ornl.gov}
}

\maketitle

\begin{abstract}
In this work, we attempt to solve the integer-weight knapsack problem using the D-Wave 2000Q adiabatic quantum computer. The knapsack problem is a well-known NP-complete problem in computer science, with applications in economics, business, finance, etc. We attempt to solve a number of small knapsack problems whose optimal solutions are known; we find that adiabatic quantum optimization fails to produce solutions corresponding to optimal filling of the knapsack in all problem instances. We compare results obtained on the quantum hardware to the classical simulated annealing algorithm and two solvers employing a hybrid branch-and-bound algorithm. The simulated annealing algorithm also fails to produce the optimal filling of the knapsack, though solutions obtained by simulated and quantum annealing are no more similar to each other than to the correct solution. We discuss potential causes for this observed failure of quantum adiabatic optimization. 
\end{abstract}


\section{Introduction}
The knapsack problem is easily stated: given a list of objects, each with a particular weight and value, how does one fill a knapsack so as to maximize total value, while remaining under a given weight limit? Despite its apparent simplicity, the knapsack problem is an NP-complete problem \cite{npcomplete}; it is among the most studied problems in combinatorial optimization, and has a number of real-world applications in economics, logistics, finance, etc. \cite{knapsack_trends, economics}. As such, many approximate algorithms exist for solving the knapsack problem \cite{knapsack_trends, approximate_algorithms, classical_knapsack}. For large instances of the knapsack problem with potential real-world applicability, these algorithms can easily fail to provide the optimal solution in a reasonable amount of time \cite{knapsack}. 

In recent years, adiabatic quantum computation has generated much interest for its potential to solve difficult optimization problems, and in particular NP-complete problems. For some instances of NP-complete problems, adiabatic quantum optimization has been able to provide a solution in only polynomial time \cite{polynomial_time}. Understandably, such prospects for solving NP-complete problems are exciting and have generated a great deal of interest in adiabatic quantum computation \cite{lucas}. However, other works \cite{boris8, boris9, boris10, boris11, boris13} have shown that in some particularly hard instances, adiabatic quantum computation actually fails to solve NP-complete problems. Theoretical work has also noted that adiabatic quantum computing fails to solve not just difficult instances of NP-complete problems, but random instances as well \cite{boris_main, boris_anderson}. It is of great relevance, then, to probe the limits of adiabatic quantum computing using current noisy intermediate-scale hardware. Theoretical work has yet to define strict bounds on the limits of adiabatic quantum computing when it comes to solving NP-complete problems.

In this work, we attempt to solve the knapsack problem on the D-Wave 2000Q adiabatic quantum computer, using a number of small problem instances (whose optimal solutions are already known, and in fact easily found using classical methods). We find that for all test problems, the D-Wave machine fails to return a lowest-energy solution corresponding to the optimal filling of our knapsack problem. We observe that the classical simulated annealing algorithm fails to solve the knapsack problem as well. However, we do not find much similarity between the solutions presented by quantum and simulated annealing, though both are non-optimal.

\section{Background \& Related Work}

Adiabatic quantum computation relies on the premises of the Adiabatic Theorem of quantum mechanics \cite{sakurai}. In practice, we choose an initial hamiltonian $H_i$, whose ground state is easy to prepare, and evolve the system in time according to
\begin{equation}
    \label{eq:adiabatic}
    H(t) = \left( 1-\frac{t}{T}\right)H_i + \frac{t}{T}H_f
\end{equation}
for $0 \le t \le T$, where $H_f$ is the final hamiltonian of the system, whose ground state encodes the solution to the optimization problem we wish to solve. The Adiabatic Theorem says that if we evolve the system slowly enough, then the system remains in the ground state throughout, and we can measure the ground state at $t=T$, thereby finding the solution to our problem \cite{annealing}. 

Using the D-Wave 2000Q adiabatic quantum computer, one can encode a problem in the form of a quadratic unconstrained binary optimization (QUBO) problem. Problems of this kind are equivalent to an Ising spin glass problem, which is itself NP-complete \cite{isingNP}. When solving a QUBO problem, we seek to maximize (or minimize) the quantity
\begin{equation}
    \label{eq:qubo}
    \bm x^T Q \bm x + \bm b^T \bm x \, .
\end{equation}
Here, $Q \in \mathbb R^{n \times n}$ and $\bm b \in \mathbb R^{n}$ are problem constants, and we seek a binary vector $\bm x \in \mathbb B^n$ which maximizes (minimizes) the above expression. Here $\mathbb B = \{0,1\}$ denotes the binary set, as it will throughout the rest of this work. So to solve the knapsack problem using the D-Wave machine, we need only formulate the knapsack problem as a QUBO problem.

\subsection{Problem Formulation}

In this section, we formally define the integer-weight knapsack problems which are the subject of this paper. We then present a formulation of the knapsack problem as a QUBO problem, so that it may be embedded on the quantum hardware.

First, we define the knapsack problem. Let $v_1,...,v_n$ denote the values of objects $1,...,N$, and $w_1,...,w_N$ their weights. For each object $i$ we may choose from, let us define $x_i \in \mathbb B$ such that $x_i=1$ if object $i$ is included in the knapsack, and $x_i=0$ otherwise. For a knapsack of capacity $W$, we seek to maximize the quantity
\begin{equation}
    \label{eq:KP}
    \sum_{i=1}^N v_i x_i = \mathcal V,\quad \text{ subject to } \quad \sum_{i=1}^N w_i x_i \le W\, .
\end{equation}
We use the same notation throughout this work to specify the weights, values, and capacity of a general knapsack problem. In order to embed a problem on the quantum hardware, we use the hamiltonian formulated by Lucas in \cite{lucas}. We restate the hamiltonian here. Let us introduce another set of binary variables $y_1,...,y_W$, where $y_j=1$ if the knapsack has total weight $j$, and $y_j=0$ otherwise. The problem hamiltonian takes the form $H = H_A+H_B$, where 
\begin{equation}
 H_A = A \left(1-\sum_{j=1}^W y_j\right)^2+A\left(\sum_{j=1}^W jy_j - \sum_{i=1}^N w_i x_i \right)^2
 \label{eq:lucas}
\end{equation}
 and $H_B =  - B \sum_{i=1}^N v_i x_i$. The constants $A$ and $B$ are defined such that $0 < B \max (v_i) < A$. In \eqref{eq:lucas}, the first term enforces that precisely one $y_j=1$, and the second term enforces that the choice of objects fill the knapsack to that particular weight $j$. The second part of the hamiltonian $H_B$ maximizes the overall value of the knapsack (and thereby minimizes the total energy). We do not want solutions which weakly violate $H_A$ at the expense of $H_B$ becoming more negative. The constants $A$ and $B$ are defined so that adding an addition item to the knapsack, which makes it too heavy, is not allowed. Evidently, the number of binary variables required to formulate a knapsack problem in this fashion scales as $N+W$. Because QUBO problems and the knapsack problem are both NP-complete, it is possible to transform one into the other in a polynomial number of steps \cite{karp}. This is not proven in the cases specifically mentioned in \cite{lucas}, however. 
 
It is easy to transform this hamiltonian into the form of \eqref{eq:qubo}, though it requires inundating the reader with a plethora of new vector quantities. First, we create a column vector $\bm z = (x_1,...,x_N,y_1,...,y_W)^T \in \mathbb B^{N+W}$ from our problem variables. We also define the vectors
\begin{equation}
    \label{eq:vectors}
    \bm W = \begin{bmatrix} -w_1 \\ \vdots \\ -w_N \\ 1 \\ \vdots \\ W \end{bmatrix}, 
    \quad \bm \lambda = \begin{bmatrix} 0 \\ \vdots \\ 0 \\ 1 \\ \vdots \\ 1 \end{bmatrix},
    \quad {\rm and} \quad \bm V = \begin{bmatrix}-v_1 \\ \vdots \\ -v_N \\ 0 \\ \vdots \\ 0 \end{bmatrix}.
\end{equation}
It is then clear to see that our hamiltonian can be written as 
\begin{equation}
    \label{eq:quboform}
    H = \bm z^T(A\bm W \bm W^T + A \bm \lambda \bm \lambda^T) \bm z + (2A\bm \lambda^T+ B \bm V^T) \bm z\,.
\end{equation}
We have neglected to include constant terms in \eqref{eq:quboform} since they do not affect the optimal choice of $\bm z$. Now we have clearly converted \eqref{eq:lucas} into the form of a QUBO problem as in \eqref{eq:qubo}, so we are ready to solve the problem using the adiabatic quantum computer. 

\section{Methods}

\subsection{Quantum Hardware}

In this work, we performed all quantum annealing tests using the D-Wave 2000Q adiabatic quantum computer. The D-Wave machine has approximately 2048 functional superconducting qubits, with roughly 5,600 inter-qubit connections. However, one can only accommodate optimization problems with a maximum of 64 binary variables, due to limitations imposed by inter-qubit connectivity. This fact implies that we are limited to solving knapsack problems with $(N+W) \le 64$. When performing quantum annealing, it is necessary to measure the ground state of the system numerous times in order to gain a probabilistic description of the system. In this work, we measure the ground state of the system between 100 and 5,000 times. 

Qubits on the D-Wave hardware are connected in a chimera graph structure \cite{pras}, and it is necessary specify a graph embedding (or mapping) to the quantum hardware in order to solve a given problem. In this work, we performed simulations using two embedding methods: the automatic embedding performed by D-Wave, and the \texttt{find embedding} method provided by the \texttt{minorminer} package from D-Wave \cite{minorminer}. 

\subsection{Classical Solvers}
\label{sec:solver}

We compare the results obtained by the quantum annealer to three classical methods for solving the knapsack problem. In this work, we solved all problem instances using the Gurobi Optimization software and a dedicated solver for the integer-weight knapsack problem from Google OR-Tools. Both of these solvers employ hybrid branch-and-bound algorithms; for more details, see \cite{branch}. We performed all classical computations on a machine with a 2.3 GHz Intel i5 processor with 4 cores, and 8 GB of memory. 

Another worthy point of comparison to quantum annealing is the classical simulated annealing algorithm \cite{simulated_algo}. We attempted to solve each knapsack problem instance presented in this work with the simulated annealing algorithm as well. 

\subsection{Problem Instances}

In order to test the abilities of adiabatic quantum computing, we attempted to solve a variety of small knapsack problems, subject to the aforementioned constraint $(N+W) \le 64$. We selected suitably small problems from the online databases \cite{database1} and \cite{database2}, whose optimal fillings were known. These problems ranged in size between 14 and 57 binary variables. Table \ref{tab:problems} provides an overview of the problem data for those problems used from \cite{database1} and \cite{database2}. In practice, these problems are trivially small, and it is possible to solve all of them using a brute-force method of simply checking all possible solutions using the hardware setup described in \ref{sec:solver}. 

We also generated instances of small knapsack problems with randomly chosen weights, values, and capacities, subject to the restriction $(N+W) \le 64$. For instance, such a problem might have weights chosen from a uniform distribution of integers between 5 and 20, values chosen from a uniform distribution of integers between 20 and 60, and capacity $W$ fixed such that $(N+W) \le 64$. We generated such problems requiring between 15 and 60 binary variables to encode on the quantum hardware.

\begin{table}[t]
\caption{Knapsack Problem Data}
\label{tab:problems}
\begin{center}
\begin{tabular}{|c|c|c|c|}
\hline
\textbf{Name} & \textbf{\textit{Number of objects ($N$)}}& \textbf{\textit{Capacity ($W$)}}& \textbf{\textit{Binary Variables}} \\
\hline
A &4 &11 &15  \\
\hline
B &4 &20 &24 \\
\hline
C &5 &26 &31 \\
\hline
D &7 &50 &57 \\
\hline

\end{tabular}
\label{tab1}
\end{center}
\end{table}

\section{Results}
\label{sec:results}
We summarize our main findings here. For the small instances of knapsack problems noted in table \ref{tab:problems}, we found that adiabatic quantum optimization fails to return the solution corresponding to optimal filling of the knapsack. Interestingly, the simulated annealing algorithm also fails to find the optimal filling in all cases, though there is no correlation between solutions returned by simulated and quantum annealing methods. Recall that the solution to a knapsack problem can be represented by a binary vector $\bm x \in \mathbb B^{N+W}$; in order to quantify the similarity between two solutions $\bm a$ and $\bm b$, we use the Hamming distance $d_H(\bm a, \bm b)$ defined by 
\begin{equation}
    \label{eq:hamming}
    d_H (\bm a, \bm b) = \frac{\sum_{i=0}^{N+W}|a_i - b_i|}{N+W}\, ,
\end{equation}
where $\bm a, \bm b \in \mathbb B^{N+W}$. Due to the probabilistic nature of quantum systems, we must perform multiple measurements to gain a clear picture of the ground state of the final hamiltonian. We shall use the Hamming distance as a measure of similarity between two solutions (in the form of binary vectors) to a given knapsack problem, averaged over a number of trials. 

In comparison to simulated and quantum annealing, the Gurobi and Google OR-Tools solvers consistently found the optimal filling for each knapsack problem attempted in this work. This result is hardly surprising, since the knapsack problems in table \ref{tab:problems} are trivially small, and can be verified by a brute-force approach which simply checks all possible solutions. Such a method was also employed in order to verify the correctness of the hamiltonian \eqref{eq:quboform}. Using knapsack problems with known optimal solutions, specifically those in table \ref{tab:problems}, we performed a brute-force search for the lowest energy solution and its corresponding solution vector $\bm x$. We found that \eqref{eq:quboform} does indeed encode the correct ground state solution to a given knapsack problem. This method of verifying the correctness of \eqref{eq:quboform} is hardly pedagogical, though in \cite{lucas} the author does not provide proof of the correctness of \eqref{eq:lucas}. However, our brute-force method applied to small problems suggests that the solution to a knapsack problem, $\bm x$, minimizes the total energy of the hamiltonian \eqref{eq:quboform} \footnote{In fact, it is not guaranteed that a ground-state solution to a given knapsack problem even exists. For more discussion see \cite{lucas}.}. 

\begin{figure}[t]
  \centering
    \includegraphics[width=0.45\textwidth]{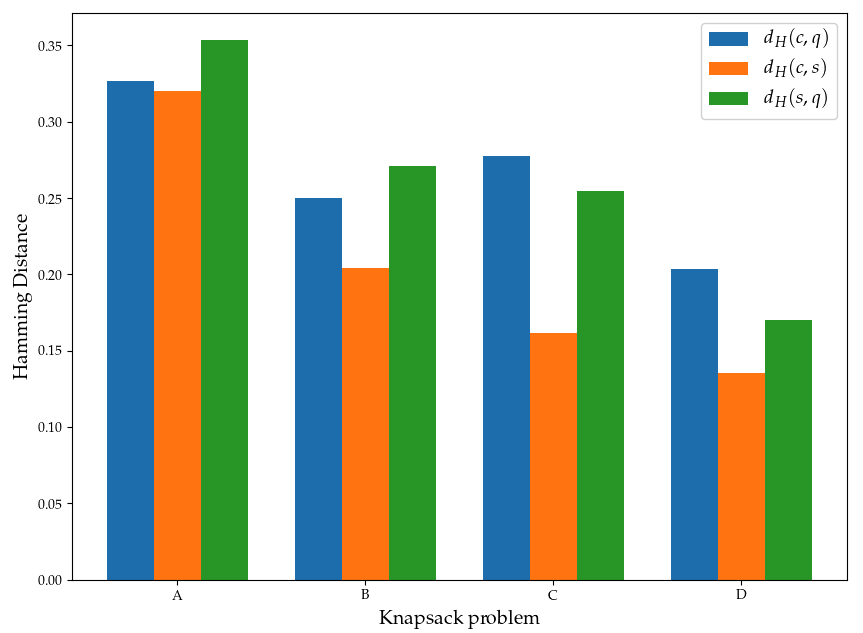}
  \caption{Hamming distances vs. problem size for knapsack problems $A$, $B$, $C$, and $D$ mentioned in table \ref{tab:problems}. Here, $d_H(\bm c, \bm q)$ stands for the Hamming distance between correct and quantum solutions, $d_H(\bm c, \bm s)$ for correct and simulated annealing solutions, and $d_H(\bm s, \bm q)$ for simulated and quantum annealing.}
  \label{fig:hamming}
\end{figure}

Now, we discuss in greater detail the results obtained using the D-Wave 2000Q and the simulated annealing algorithm. Both methods for solving the knapsack problem are inherently probabilistic \cite{simulated_algo}; in order to measure the similarity between solutions to a given knapsack problem, we use the Hamming distance introduced in \eqref{eq:hamming}. We denote solutions obtained on the D-Wave machine by $\bm q \in \mathbb B^{N+W}$, those on simulated annealing by $\bm s \in \mathbb B^{N+W}$, and the correct solution as $\bm c \in \mathbb B^{N+W}$. Hamming distance versus problem size is presented in figure \ref{fig:hamming}. For results shown in this figure, we performed 100 runs on the quantum processor to calculate a minimum energy solution $\bm q$, and then calculated the average Hamming distance for 10 separate calculation of $\bm q$. We used the automatic embedding method provided by D-Wave to perform these calculations. We also performed the same same tests with only one calculation of $\bm q$, and also noticed a trend similar to that in figure \ref{fig:hamming}. 

Interestingly, we observe that the Hamming distance decreases as the size of the problem increases. One can also see that simulated annealing tends to produce solutions which are closest to the correct solution as measured by $d_H(\bm c, \bm s)$. It does not appear that for a given problem, simulated and quantum annealing produce similar solutions. In fact, as the problem size increases, it appears that simulated annealing produces solutions closest to the correct solution, or that $d_H(\bm c, \bm s)$ is smaller than both $d_H(\bm c, \bm q)$ and $d_H(\bm s, \bm q)$. For totally uncorrelated results $\bm a, \bm b \in \mathbb B^{N+W}$, one might expect that $d_H(\bm a, \bm b) \sim 0.5$. On the other hand, our results indicate that $d_H(\bm a, \bm b)$ decreases as problem size increases. However, this could simply be a consequence of the quantum and simulated annealing algorithms becoming stuck in a local minimum. This explanation is fleshed out in more detail in section \ref{sec:discussion}.

We also studied the effects of the magnitude of constants $A$ and $B$ for each knapsack problem. We tested three approximate regimes, where $A \gtrsim B \max (v_i)$ (implying that $A$ and $B\max (v_i)$ are of the same magnitude, though $A$ the larger), $A > B\max (v_i)$, and $A \gg B\max (v_i)$. For instance, we tested $A = B\max(v_i)+2$, $A = 2B\max (v_i)$, and $A=100B\max(v_i)$. For each of these regimes, we also varied the magnitude of $B$ between $1 \le B \le 100$. 

In our tests, we observed that the Hamming distance was essentially independent of a choice of $A$ and $B$, for each of the three regimes mentioned above. We tested all problems listed in table \ref{tab:problems}, and note that this result is consistent across all problem sizes. Obviously the ground state energy of our hamiltonian is impacted by our choice of $A$ and $B$; however, we did not observe minimum energy solutions returned by the D-Wave machine which were closer to the correct solution as a result of varying $A$ and $B$. 

Additionally, we varied the number of times the state of the quantum machine was sampled, in order to gain a perhaps more accurate probabilistic description of the system measured at $H_f$. As mentioned previously, the solution $\bm q$ is that which is of the lowest energy, out of all samples of the quantum system. In our work, we performed 100, 500, 1000, and 5000 samples of the quantum state of the system, in case the probability of observing the true ground state of the system is exceptionally small. Despite performing more samples, we failed to observe any solution $\bm q$ corresponding to the solution to a given knapsack problem. This result was true for all problems we tested listed in table \ref{tab:problems}. We did notice that the degeneracy of minimum-energy solutions returned by the quantum system was typically high; for instance, in a run in which the system was measured 100 times, perhaps 80 unique solutions were observed, many of which were degenerate in energy. This result was generally consistent even when we increased the number of measurement of the ground state to 5,000.

Finally, we discuss the impact of an embedding method these specific knapsack problems on the quantum hardware. The two embedding methods employed in this work seemed to require a similar number of qubits. We compared the solutions returned for a particular knapsack problem by the quantum machine for both embedding methods (with all other problem details the same). We observed that the two embedding methods did not always supply the same minimum-energy solution. This is to be expected, since quantum systems are inherently probabilistic. Typically, if $\bm q_1$ and $\bm q_2$ represent two solutions returned by different embedding methods, we observed that $d_H(\bm q_1, \bm q_2)$ was small on average. This evidence suggests that the two embedding methods produce quite similar results when all other problem variables are held constant.

\section{Discussion}
\label{sec:discussion}
From the results presented in section \ref{sec:results}, it is clear that adiabatic quantum optimization fails to produce the solution to many knapsack problems. Where, then, does the method fail? This turns out to be a subtle question for a few reasons. In the process of adiabatically varying the system from the initial to final hamiltonian, it is impossible to track the evolution each energy eigenstate, and the presence of exponentially small energy gaps $\Delta_i$ between eigenstates or other issues that may arise as the system evolves. As a consequence, adiabatic quantum computers may get stuck in local minima as the algorithm progresses, and fail to return the true ground state solution \cite{boris_main, boris_anderson}. Reference \cite{boris_main} describes the failure of adiabatic quantum optimization for random instances of NP-complete problems, which arise due to exponentially small gaps in the spectrum of the final hamiltonian. By the Adiabatic Theorem, if the gaps in the spectrum evolve as $\Delta_i(t)$, then the overall computation time scales as $T \sim 1/\Delta_{\rm min}^2$, roughly speaking; exponentially small gaps $\Delta_i$ could lead to exponentially long $T$ required to solve the problem.

Previous work \cite{polynomial_time} has shown that some small instances of NP-complete problems can be solved in polynomial time through adiabatic quantum optimization. However, other analytic results suggest that exponentially small gaps in the spectrum of the 3-SAT problem, another well-known NP-complete problem, occur in certain instances \cite{boris10, boris13, boris14}. The phenomenon of Anderson localization has also been shown to make adiabatic quantum optimization fail \cite{boris_anderson}. To be certain, the exact causes for the failure of adiabatic quantum optimization in this work are impossible to probe directly. Even with direct access to the inner workings of the D-Wave machine, it is probably impossible to continuously measure and evaluate the adiabatic evolution of a hamiltonian $H(t)$. 

One interesting point we noted in figure \ref{fig:hamming} was the decrease in the Hamming distance $d_H(\bm a, \bm b)$ as the problem size increased. It is likely incorrect to interpret this result to mean that simulated and quantum annealing come closer to the correct solution $\bm c$ as problem size increases. If $\bm q$ and $\bm s$ are both stuck in local minima, for instance, then we could expect $d_H(\bm q, \bm c), d_H(\bm s, \bm c) \le 0.5$ naively. In other words, solutions in local minima to "look" more like the correct solution $\bm c$, at least in terms of the Hamming distance. In the future, when larger, more robust quantum hardware is available, we will be able to study problems requiring more than 64 binary variables. It will be interesting to see whether Hamming distances are smaller for far larger problems, though this result probably does not imply increased correctness of simulated and quantum annealing. This could be interpreted as a tendency for quantum and simulated annealing to become stuck in local minima.

\section{Conclusion}

Adiabatic quantum computing is a promising technology which has demonstrated its ability to solve certain optimization problems which are difficult for classical computers \cite{polynomial_time, quboformulations}. As a consequence, many NP-complete problems have been formulated for adiabatic quantum computing \cite{lucas} despite the novelty of the platform, its inherent noisiness, and limited size \cite{preskill}. In this work, we demonstrate limitations of current-generation quantum hardware in the noisy intermediate-scale quantum (NISQ) era, as applied to solving the NP-complete knapsack problem. We find that adiabatic quantum optimization is unable to provide the optimal solution to a variety of small knapsack problems, within the limitations imposed by currently available hardware. While this results is discouraging for the prospects of solving certain NP-complete problems with real-world applicability, it is nearly impossible to point to a cause of the failure of AQC as demonstrated in this work. It is possible that in the future, when large-scale, robust quantum hardware is perhaps available, one may be able to solve the knapsack problems presented in this work with great efficiency. However, due to the subtlety of state evolution during the process of adiabatic optimization, it may ultimately be impossible to solve such instances of the knapsack problem. This could be due to exponentially small gaps in the spectrum of the time-evolved hamiltonian $H(t)$; in this case, computing the solution to a given knapsack problem would take an exponential amount of time \cite{boris_main}. We hope to answer such questions in the future when better hardware is presumably available. Hopefully, the fundamental limits of adiabatic quantum optimization can be better understood by this undertaking.

\section*{Acknowledgment}
We thank D-Wave Systems for access to their 2000Q adiabatic quantum computer, which was used in this work. 

\section*{Conflict of Interest}
The authors declare no competing interests.


\vspace{12pt}

\end{document}